# Bibliometric Profile of an Emerging Journal: Participatory Educational Research

Rumiye Arslan
*Department of Elementary Education, Faculty of Education, University of Amasya, Amasya, Turkey ORCID ID: 0000-0002-6679-294X*

Keziban Orbay
*Department of Mathematics and Science Education, Faculty of Education, University of Amasya, Amasya, Turkey ORCID ID: 0000-0002-7642-4139*

Metin Orbay*
*Department of Mathematics and Science Education, Faculty of Education, University of Amasya, Amasya, Turkey ORCID ID: 0000-0001-5405-2883*

"Participatory Educational Research (PER)" journal is one of the journals that contributes to the field of education and indexed in major international databases such as ERIC and Scopus. This study provides the bibliometric characteristic of the total 347 articles published in PER during the period of 2014-2021 using bibliometric analysis. Publish or Perish software to collect citation data from Google Scholar was used as an analysis instrument for the impact of the articles. It was found that short-titled articles received more citations than long-titled articles (over 2 times greater), but not statistically significant ($p>0.05$). On the other hand, correlation between citation and download numbers was found to be a statistically significant positive ($r_S=0.289$ and $r_P=0.524$; $p<0.01$). In the analysis of keywords and titles, it was observed that the prominent words overlapped with each other and with the purpose of journal as well. The most cited articles and the institutions contributing to national and international levels were analyzed too. It was concluded that 83.72% of the authors were in Turkey, there was no "institutional localization" among the institutions contributing at the national level and that they had achieved significant success in terms of national recognition. PER has gained significant momentum in academic standards and visibility since it first joined the umbrella organization DergiPark in 2019. It should be noted that amongst the most important points toward being open to development in the point of international recognition is the existence of contributions from Anglo-Saxon and Continental European countries, which have appeared as limited. According to the findings, it is discussed what can be done from this point onward based on basic publishing standards, publication content, national/international visibility, and citation analyses. The results can guide authors during the writing phase of studies and the editors and referees during the selection and evaluation phases.

* Correspondency: metin.orbay@amasya.edu.tr

## Introduction

Knowledge from past to present has been the most important source of development of societies and has brought different societies together through various communication channels. The scientific accumulation that emerges through knowledge is shared through communication channels and makes it possible to discuss different ideas in the scientific literature. In particular, books and academic journals play a leading role among the official communication languages of science in the construction, dissemination and use of information (Abramo, 2018; Hicks, 2012; Riviera, 2013; Sivertsen & Larsen, 2012). However, the communication sources used between the branches of science may differ over time. The main reasons for this difference include many reasons such as whether the area has universal or local characteristics, is up-to-date or retrospective, and is suitable for teamwork or individual work (Andersen, 2000; Diem & Wolter, 2013; Hicks, 1999; Hicks, 2012; Nederhof, 2006; Uçak Özenç & Al, 2008). The production and consumption of information is largely carried out at the national level for many reasons such as the fact that social sciences are not as biased and universal as medicine, engineering and basic sciences, and that the information produced is affected by regional, personal and social conditions, political and cultural climate (Andersen, 2000; Diem & Wolter, 2013; Hicks, 1999; Moed, 2005; Uçak Özenç & Al, 2008). Therefore, while the concern for universality and currency features academic journals in the field of medicine, engineering or basic sciences as a primary communication tool, the absence of time pressure in most sub-sciences of social sciences, especially the importance of retrospective and archival knowledge, makes books stand out as a communication tool rather than academic journals (Hicks, 2012; Uçak Özenç & Al, 2008).

The field of education is a sub-science field of social sciences where academic journals have come to the forefront in recent times (Aman & Botte, 2017; Orbay, Karamustafaoğlu, & Miranda, 2021; Örnek, Miranda, & Orbay, 2021). It is apparent that academic journals have started to come to the forefront as a widely used communication tool due to internationalization in the field of education, rapid development of field relations with other branches of science over time and developments in information technologies (Engels et al., 2012; Larivière et al., 2006; Rowlinson et al., 2015; Henriksen, 2016). As Goodyear et al. (2009) emphasized, this field is inherently interdisciplinary within social sciences and is very open to interaction with other disciplines due to its wide content. Budd and Magnuson (2010), in their studies examining the citation patterns of articles published in journals that publish heavily on higher education, have revealed that journals were cited by 45.5%, while books were cited with a rate of 26.3%. Similarly, in a study of educational research by the twenty most prolific European countries between 2002 and 2013, Aman and Botte (2017) revealed that the number of articles produced by international cooperation in education increased steadily from 14.1% in 2002 to 21.7% in 2013. In a similar study, Orbay et al. (2021) showed that internationalization increased from 20.6% in 2013 to 29.1% in 2018.

Today, advances in information technologies have made it possible and cheaply to access information easily and have increased the amount of information available by doubling every day (Fire & Guestrin, 2019). The competitive environment created by the increasing number of academic journals over time has brought along the idea of "*publish or perish!*" among researchers and also questions of "*quality or quantity?*" in studies conducted (Civera et al., 2020; McGrail, Rickard, & Jones, 2006; Van Dalen, 2021). Therefore, it is becoming more and more important to follow the publications produced in academic journals, to determine the characteristics of academic journals and publications and to analyze them based on various criteria, to understand the present and to make inferences between past and future (Engels et al., 2012; Larivière et al., 2006; Öner & Orbay, 2022; Rowlinson et al., 2015; Henriksen, 2016). Therefore, it is widely used as a quality measurement tool to analyze the environments in which scientific information is published with various mathematical and statistical techniques (Donthu et al., 2021). One of the widely used tools for measuring the contribution of journals to the field of science is Bibliometric analysis method, first defined by Pritchard (1969).

In this context, “Participatory Educational Research (PER)-ISSN: 2148-6123)” journal, which started its publication life in English in 2014 as an open access in the field of education, has become one of the journals contributing to the field by being indexed in major worldwide databases at national and international level in a short time (PER, 2022).

PER started its academic publication life as two issues per year and was published as three and four issues per year over the years. Starting in 2022, it has decided to publish the journal as six issues per year. No fees are charged to the researchers under any name in the evaluation or publication processes of articles. Studies are subjected to peer evaluation by double-blind peer-review method. It is designed as a journal that is open to research carried out by qualitative, quantitative, or mixed methods in all areas of education. The main objective of the journal is to improve the quality of learning, teacher, and teaching by contributing to the educational processes carried out at all levels of education starting from preschool education, and to contribute to the improvement of their processes and results (PER, 2022). In 2019, PER has started to carry out its publication life under the umbrella of DergiPark platform in TUBITAK ULAKBIM, which provides electronic hosting and editorial process management services for academic peer-reviewed journals published in Turkey (ULAKBIM, 2022). The journal has started to be indexed in Scopus and ERIC (The Education Resource Information Center) databases, where reputable journals in the field of education at international level are indexed. On the other hand, the ESCI (Emerging Sources Citation Index) application in the Web of Science (WoS-Core Collection by Clarivate Analytics) database is under evaluation (PER, 2022).

### *Aim of the Study*

In this study, it was aimed to analyze articles published in PER during the period of 2014-2021 using bibliometric analysis, which is one of the qualitative research methods**.** For this purpose, answers to the following research questions (RQ) were sought:

**RQ1:** How do the number of articles and the number of authors change by year? What is the distribution of authors according to the institutions they work for at national and international levels?

**RQ2:** How are the articles distributed by author and citation numbers? What are the most cited articles?

**RQ3:** What are the words commonly used in the titles and keywords of articles? Is there a correlation between the number of words in the title and the number of citations received by the article?

**RQ4:** Is there a correlation between the number of downloads of articles and the number of citations they receive?

## Methodology

The sample of this research consists of 347 articles published in PER between 2014 and 2021. The identity information and content analyses of the articles were carried out with the data obtained via DergiPark platform (ULAKBİM, 2022). Publish or Perish software (Harzing, 2007) was used to determine the citation numbers. In this software, the Google Scholar index, which is considered the index with the most common scanning network and scans not only printed articles but also many early print and institute stores, has been selected (Doğan, Şencan, & Tonta, 2016; Harzing & Alakangas, 2016; Jacsó, 2005; Jacsó, 2009). Duplicate references to related articles, references from sources other than articles, books and papers have been cleared. The citation scan was conducted for the period from January 20, 2022, to January 25, 2022. Throughout the study, the focus was on articles and review studies. However, since only one review study was detected in the specified period, the article will be named jointly for the reviews and articles in the following sections.

Whether the obtained data was parametric or not was determined as a result of the analysis (George & Mallery, 2010). According to the Kolmogorov-Smirnov Test and the descriptive statistics of variables results (skewness, kurtosis values and graphical representations such as histograms, Q-Q plots and box plots), as the data found to be skewed, the data were presented as medians and interquartile ranges. And then the comparisons between the number of total citations, the number of words in the title and the visibility were performed using the nonparametric Mann-Whitney U test for two groups and Kruskal-Wallis test for three groups. Spearman's coefficient ($r_S$) test was used to investigate the correlation between the number of words in the title, the number of downloads and the citation counts. A $p$-value less than *0.05* was considered statistically significant, and IBM SPSS 20 software (Armonk, NY, USA) was used for the analysis of data.

## Findings and Discussion

In this section, findings related to research questions were given and the results were discussed with relevant literature.

### *Findings and Discussion for RQ1*

The change in the number of articles and authors depending on the years is given in Figure 1. As shown from Figure 1, 347 articles were published during the period in question and the number of articles varies significantly depending on the years. However, in academic journalism, the number of articles in each issue of the journal is expected to be balanced both on a number/volume basis and depending on the years (Clarivate Analytics, 2022; Scopus, 2022; TR Index, 2022). Meanwhile, the average number of authors per article remains almost linearly constant.

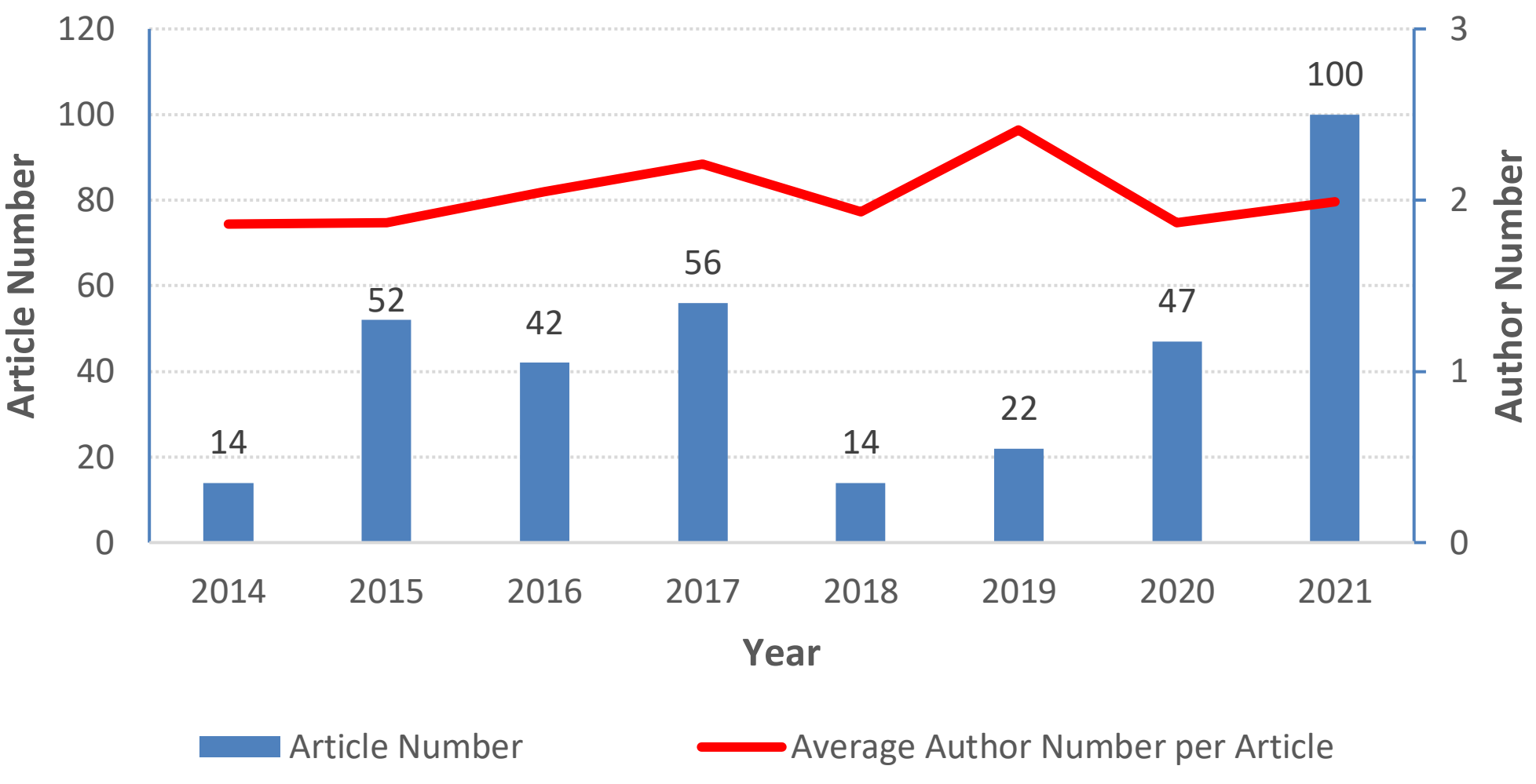

**Figure 1.** Year-wise Distribution of Articles and Author Number.

It can be considered that, among the main reasons why such a situation has arisen in the change of the number of publications in PER, two special issues were published in 2015-2016 and that the journal has published different numbers over the years. On the other hand, it can be considered that the journal has started to be indexed in ERIC and Scopus, which are accepted internationally, since 2019. This interest brings positive correlation between the qualities and quantities of the articles. One of the main reasons for this situation is that publishing in journals with high impact value indexed in important databases is considered very prestigious within the academic ecosystem and is a natural result of the supply and demand shown to these journals (Huang, 2016).

During this period, all articles received a total of 1675 citations. The majority of citations are citations from journals indexed in Scopus and ERIC, mostly TR indexes. In this context, the articles published in the period were written by 533 authors, and it was concluded that 83.72% of the authors were in Turkey. On the other hand, it was seen that authors from 19 different countries besides Turkey contributed to the journal with their articles. The most prolific ten countries based on the percentage of these authors was given in Figure 2.

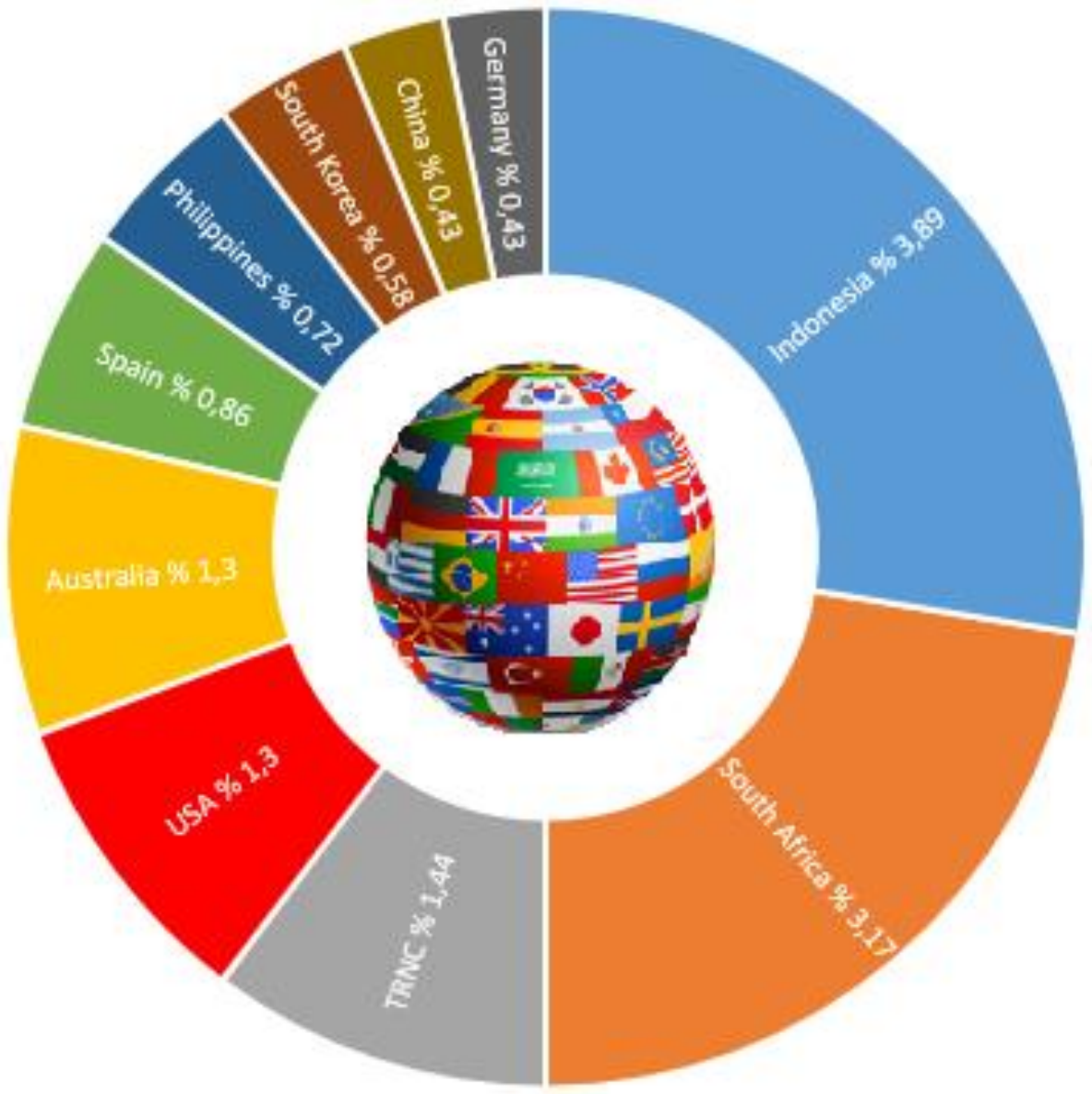

**Figure 2.** The Most Prolific Ten Countries Based on % of Total Author.

During the analysis process of the articles from Turkey, authors working in different schools affiliated with the Ministry of National Education were gathered under the umbrella of “Ministry of National Education-Ankara”. It is observed that there are authors from 99 different institutions from Turkey. On the other hand, researchers from countries other than Turkey were grouped within themselves and found to have contributions from a total of 45 different institutions. In other words, it is possible to say that articles from 144 different institutions were published in PER during the period in question. Based on this data, the distribution of the top ten contributing institutions at national and international level according to the number of authors is given in Table 1.

**Table 1.** The First Ten National and International Institutions by Total Author.

| ***Institutions-National*** | **TA %** | ***Institutions-International*** | **TA %** |
|---|---|---|---|
| Amasya University | 10.52 | South Africa University-South Africa | 2.88 |
| Ondokuz Mayıs University | 7.93 | Universitas Negeri Malang-Indonesia | 1.15 |
| The Ministry of National Education | 5.48 | Flinders University-Australia | 1.01 |
| Necmettin Erbakan University | 4.32 | Near East University-TRNC | 0.86 |
| Gazi University | 3.75 | University of Western Macedonia-Greece | 0.58 |
| Afyon Kocatepe University | 3.03 | Islamic University of Maulana-Indonesia | 0.58 |
| Karadeniz Technical University | 2.88 | Woosong University-South Korea | 0.58 |
| Ağrı İbrahim Çeçen University | 2.47 | Seville University-Spain | 0.58 |
| Gaziosmanpaşa University | 2.16 | Braunschweig University-Germany | 0.43 |
| Hacettepe University | 1.44 | Texas A&M University-USA | 0.43 |

*TA= Total Author.*

It is noticeable that the most active institutions at both national and international level are universities, and especially education faculties come to tkm .he fore. This situation is in agreement with the literature (Moed, 2006). Examining the distribution of the top ten institutions, Amasya University (10.52%), Ondokuz Mayıs University (7.93%) and The Ministry of National Education (5.48%) are in the top three. Considering the contribution rates of the top ten institutions at national level, there is no “institutional localization”. When the institutions in which the authors worked at the national level were examined on a provincial basis, it was seen that the articles were published from all provinces except only 19 provinces out of a total of 81 provinces. On the other hand, it is seen that the vast majority of provinces that do not publish articles are relatively small in population (Ardahan, Iğdır, Yalova, Bitlis, etc.) or that there is no institute or faculty of education within the university or is not active in these provinces. In this context, as seen from Figure 3, it is possible to say that PER has achieved significant recognition on the academic platform in a short period of time within the scope of Turkey.

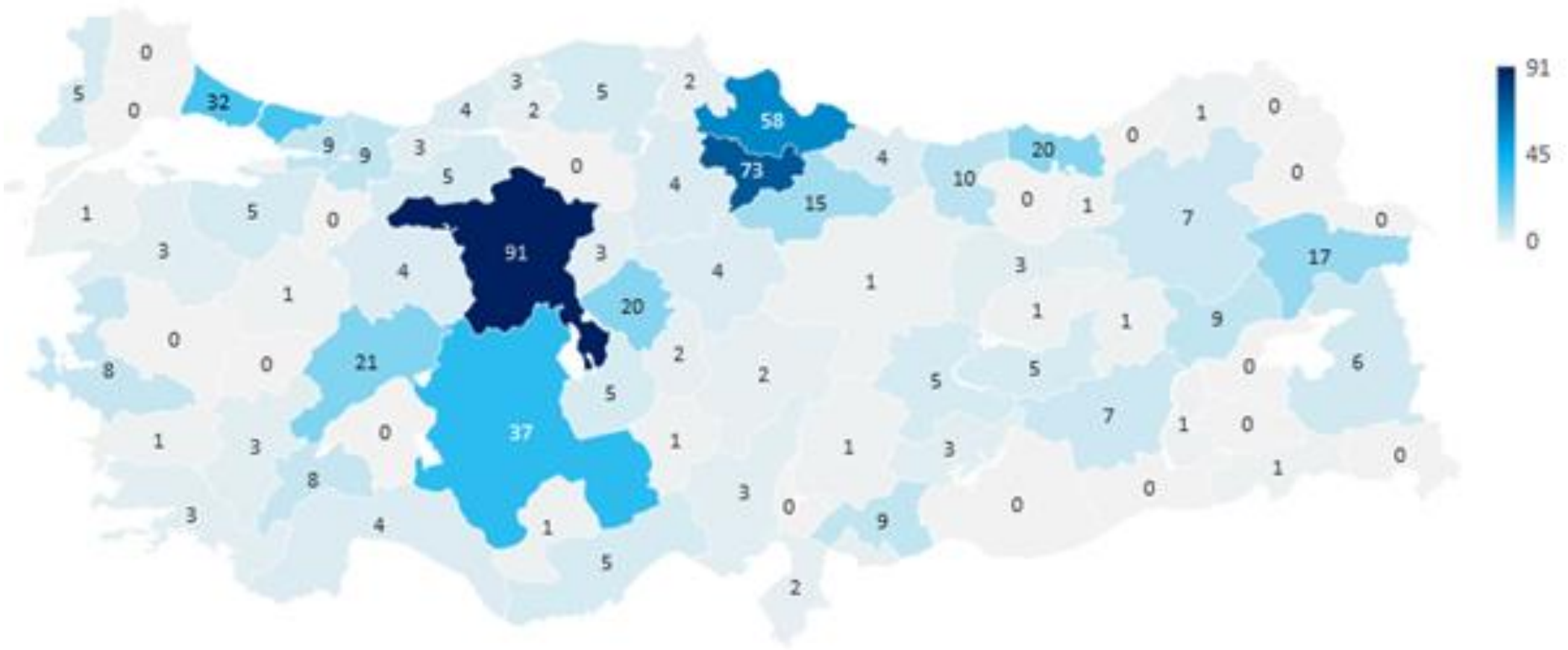

**Figure 3.** Distribution of Authors of Articles Addressed to Turkey by Province.

Among the most basic requirements of scanning in internationally accepted indexes, the authors, editors, and members of the editorial advisory board of the journal want to have an international diversity in accordance with the target audience of the journal (Clarivate Analytics, 2022; Scopus, 2022). If the content of the journal is aimed at an international audience, it is desirable that the authors with international diversity who will contribute to the goal will be visible at this point. When the PER editorial board is examined, it is seen that there are expert researchers from 16 different countries. Countries other than Turkey include researchers from Spain (7), USA (6), South Africa (5), UK (3), Canada (3) and Australia (3). Given that author participation at international level is 16.8%, it can be considered that PER at first glance fits the definition of "*Globally national-locally international journal*" (Pajić & Jevremov, 2014). This characterization has been passed into the literature as a concept used for international authors or journals that do not receive references from articles published in relatively international journals (Tonta, 2017). It is known that this applies to Turkish-addressed journals in citation indexes even before such a definition is introduced into the literature. As such, the findings of a study conducted through 71 journals addressed in Turkey in citation indexes reveal that at least one author is from Turkey in four out of every five publications in the journals examined (Doğan, Dhyi, & Al, 2018). However, at this point, it can be considered that the fact that the fields of study in a social science such as education are mostly aimed at the local audience weakens the possibility of having a comprehensive international diversity. In this case, such journals are typically expected to have a more modest citation effect and measurements can be made by looking at the authenticity of the content of the journals.

PER is indexed in "Education" and "Developmental and Educational Psychology" categories in the Scopus database. As seen from Figure 4, while the journal ranks 1088th in terms of citation among 1319 journals in the Education category, it is the 295th out of 332 journals in the "Developmental and Educational Psychology" category (Scopus, 2022). A total of 79 articles indexed between 2017 and 2020 received 31 citations during this period, and Scopus CiteScore 2020 is 0.4 (Calculated on 05 May 2021). On the other hand, the total number of citations received by the 178 articles published in PER and indexed in the Scopus database is currently 115, and in other words, CiteScore Tracker 2021 has increased to 0.6 (Last updated on 07 March 2022).

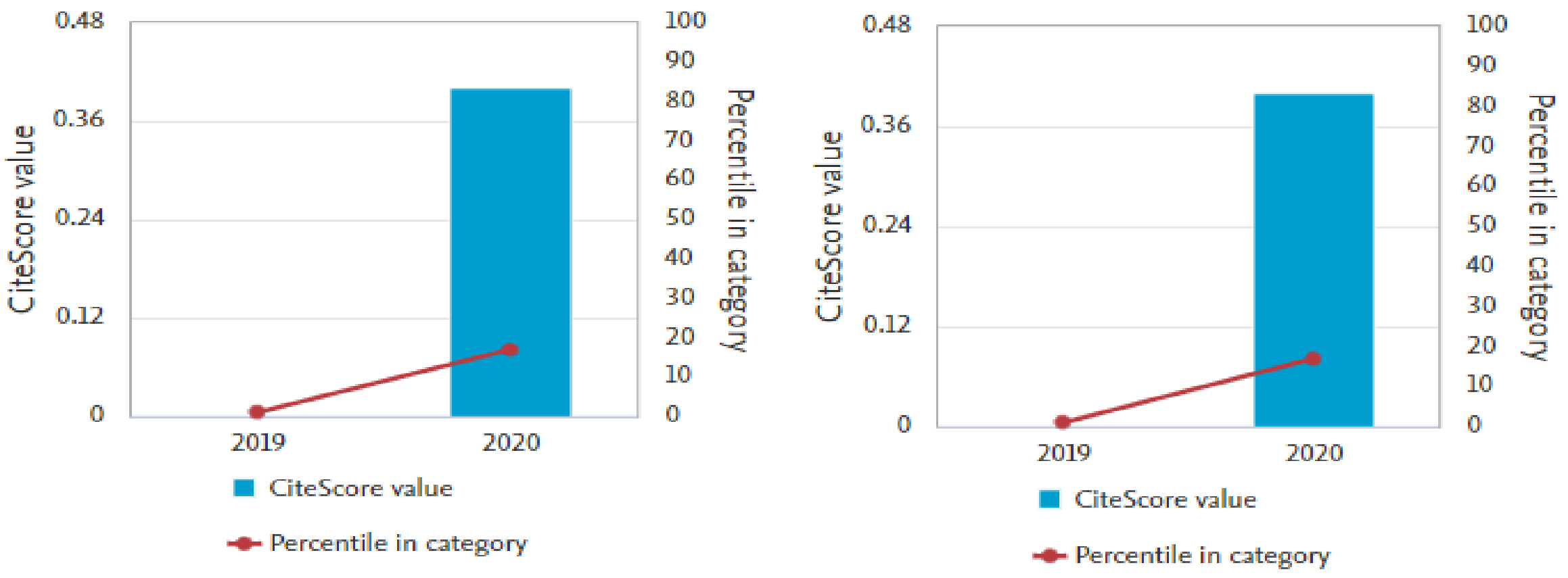

**Figure 4.** CiteScore trend for **a)** "Education" category **b)** "Developmental and Educational Psychology" category (Scopus, 2022).

On the other hand, the articles published in the journals indexed in the WoS database are considered the most prestigious in the academic community (Birkle et al., 2021; Li, Rollins, & Yan, 2018; Pranckutė, 2021). The distribution of the references received from the journals in the WoS database according to the journal quartiles (Orbay, Miranda, & Orbay, 2020) published in PER was also studied. The articles published in PER received a total of 203 citations from the journals in the WoS database. This means that 12.12% of the total citations to PER were made from journals in the WoS database. In the case of a journal in more than one quarter, the high quarter was taken into account using the optimistic approach (Liu, Hu, & Gu, 2016) and the results obtained were summarized in Figure 5.

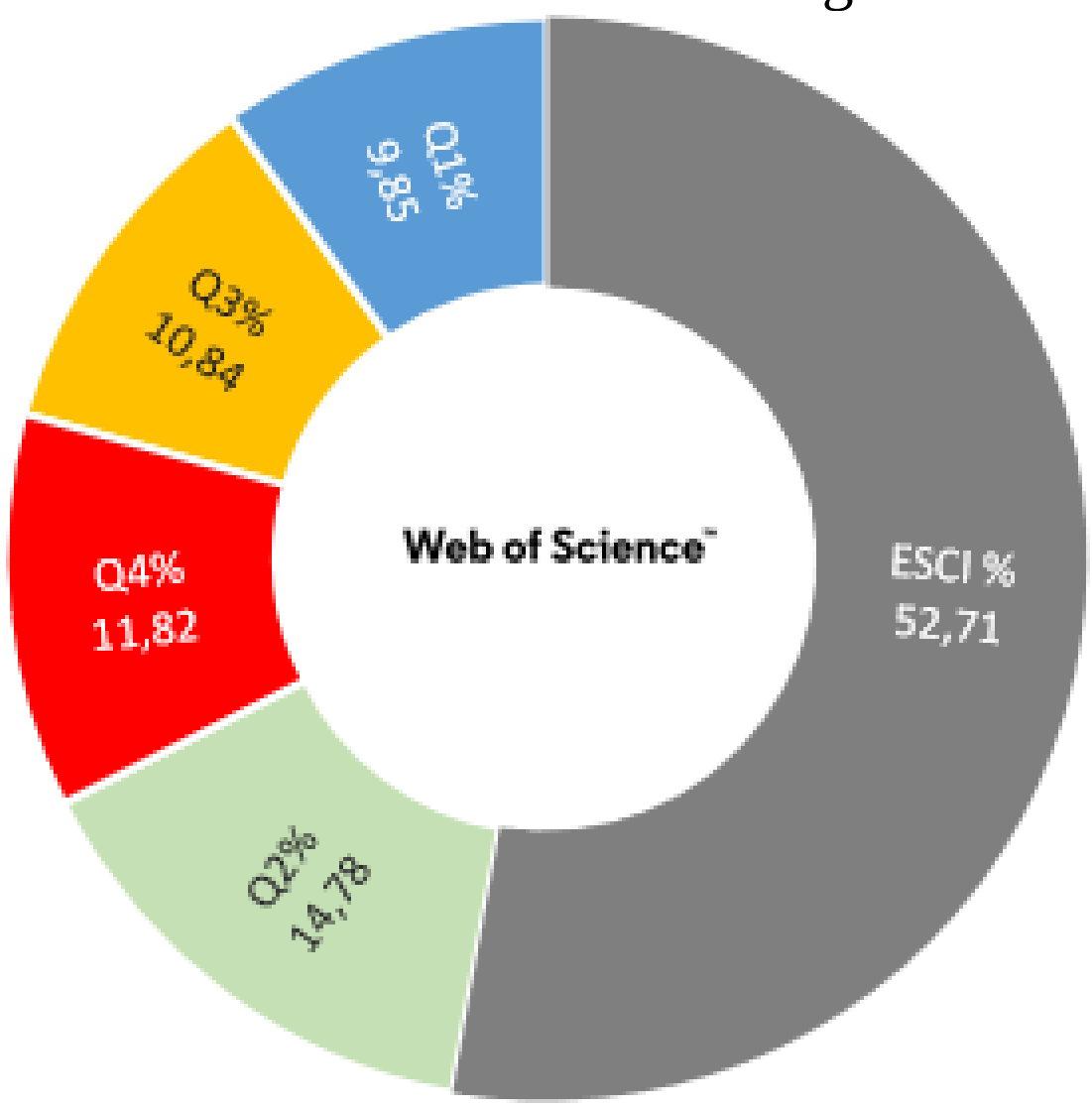

**Figure 5.** Distribution of Citations from WoS Database by Journal Quartiles.

As seen from Figure 5, 52.71% of citations from the WoS database come from journals indexed in ESCI. On the other hand, the citation rate from journals in with high impact value (Q1, Q2) is 24.63%. The rest (22.66%) comes from journals in the low quartiles (Q3, Q4).

### *Findings and Discussion for RQ2*

One scale that can be considered as an important criterion for teamwork rather than individual studies, especially in educational research, is the number of authors in the relevant research. Between 2014 and 2021, the articles published in PER were written by an average of 2.02 authors (median=2). Single and multi-authorship status percentages are given in Figure 6.

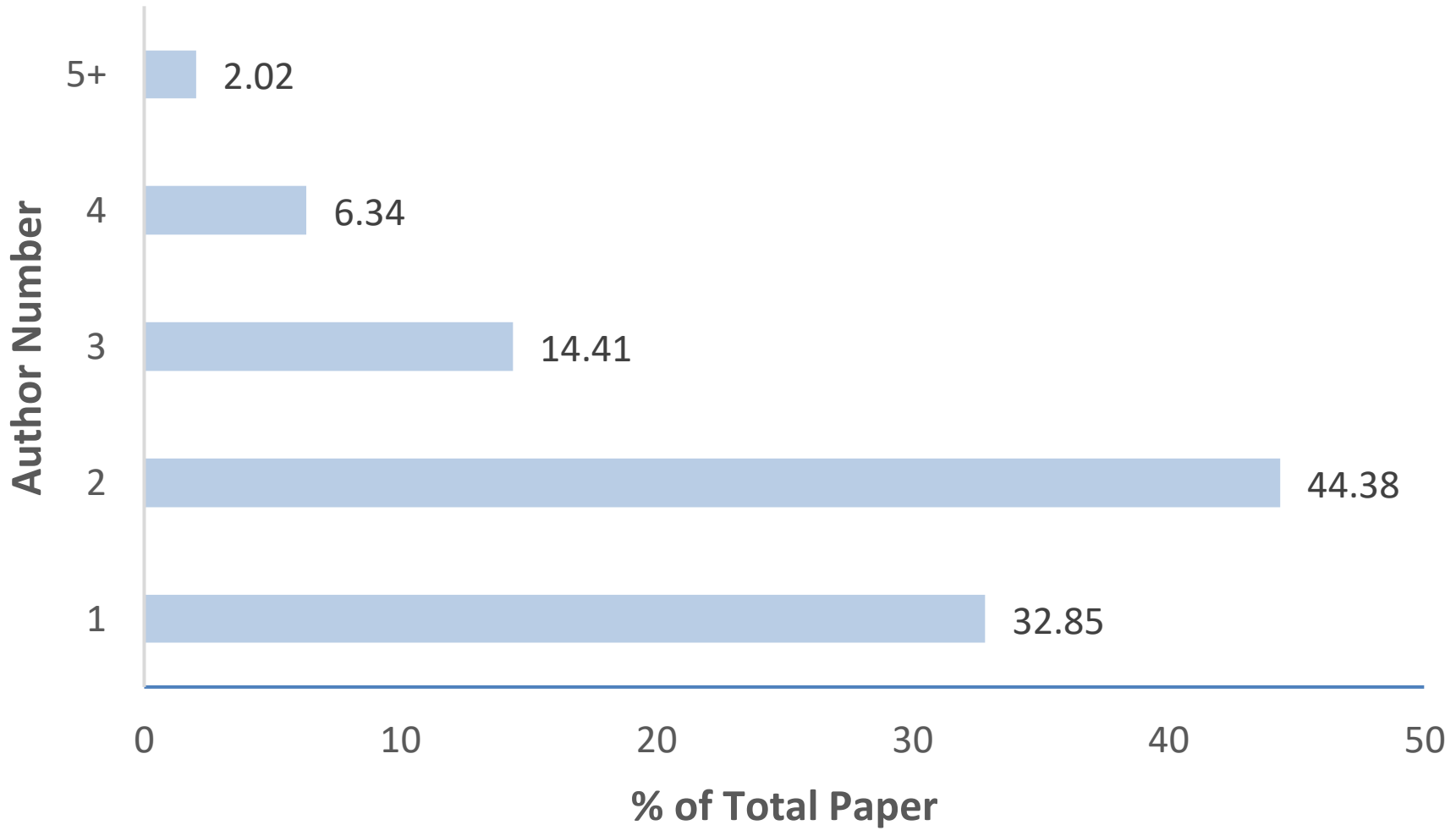

**Figure 6.** Percentages of Single and Multi-authorship Status.

Henriksen (2016) found out in his research covering the sub-sciences of social sciences that publications in educational research were written as single authors between 1980 and 2000 and with two authors between 2001 and 2013; on the other hand, found out that research in the field of special education was written with two authors between 1980 and 2010 and three as of 2011-2013. Örnek et al. (2021) emphasized that the research was written with three authors in a similar study focused on the field of special education for the period 2014-2018. Therefore, in terms of the number of authors, the findings of this study are in line with the studies in the international literature.

For the connections between the authors (a total of 89 authors) whose two or more articles were published in PER during that period, the network density map was given in Figure 7 (red=high-density; blue=low-density) using the VOSviewer 1.6.13 program (Van Eck & Waltman, 2010), which is often used for bibliometric mapping demonstration. Since the time frame in question is not very large, only a few cluster structures stand out. When articles and citations were analyzed in detail, the articles published by Korkmaz Ö, Çakır R and Kayaoğlu MN, who were among the most prolific authors (3.17, 2.30 and 1.15% of total papers), contributed over 3% to the total citations from the journal (7.40, 3.05 and 3.10% of total citations) respectively. On the other hand, the articles published by Saltan F, Ocak G and Ocak İ, who were among the most prolific authors (1.73, 1.44 and 1.15% of total papers), contributed less than 1% to the total citations from the journal (0.19, 0.18 and 0.06% of total citations). Among the most prolific authors outside Turkey was only Norma MN (2.02% of total papers and 1.43% of total citations) from South Africa.

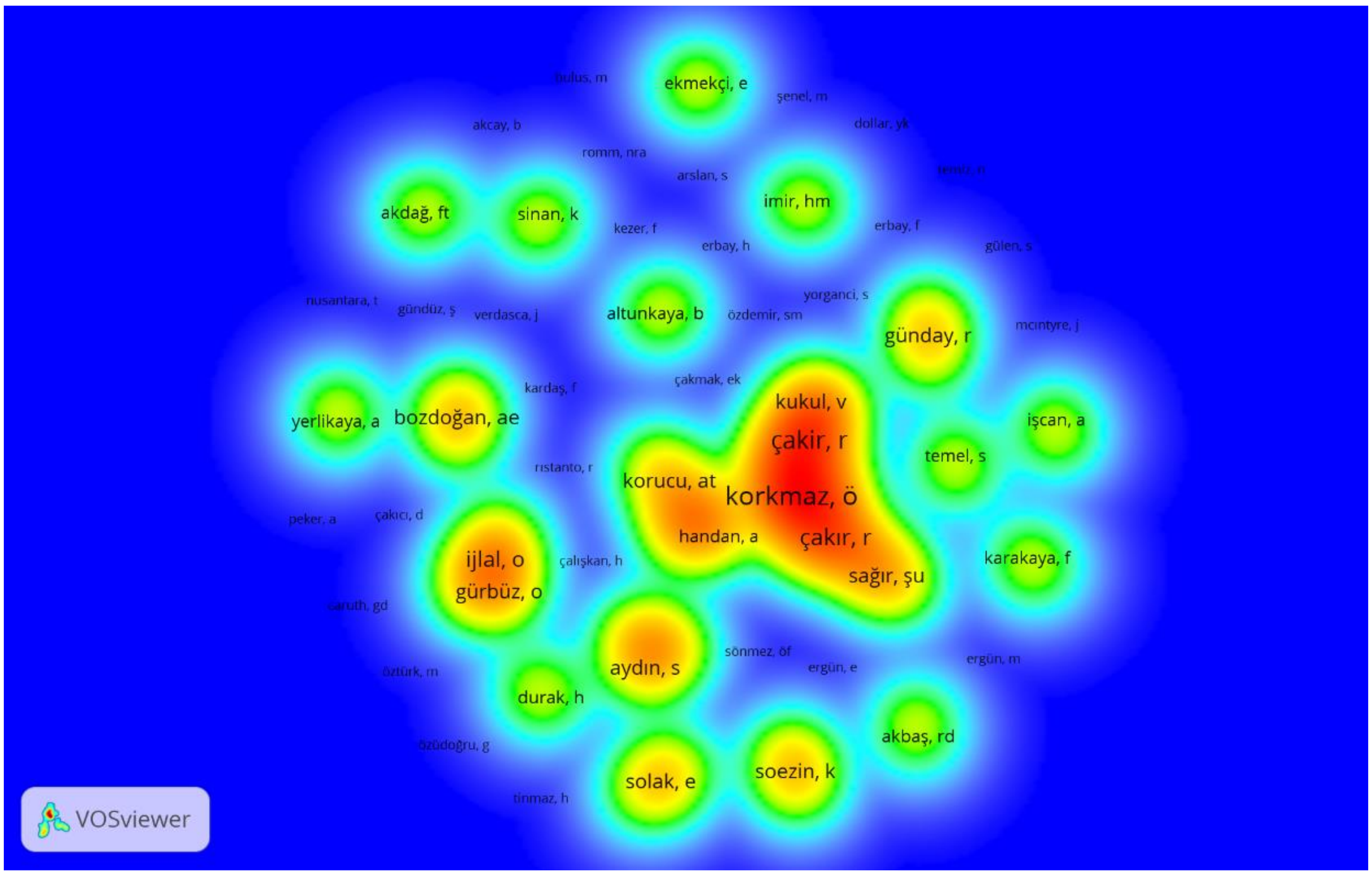

**Figure 7.** Author Co-occurrence Network Density Distribution Map.

The average number of citations per article in the specified period is 4.69 (median=1). On the other hand, the percentage of articles not yet cited is 35.73%. However, it should be noted that the percentage of articles published in the last two years is 49.20% in this rate. As expected, the share of articles with no citations is greater at the recent years as these articles had less time to accumulate citations than an older one. However, these articles still have the potential to receive citations in the future due to the nature of social sciences (Archambault &

Larivière, 2010). Although the average number of citations per article seems high, it is often not the right approach to make general judgments about all articles in the journal based on the citations that articles receive because the most important problem to encounter in the journal evaluation process is the skewness of citation, as in almost all journals (Albarrán, & Ruiz-Castillo, 2011; Bornmann, & Leydesdorff, 2017; Seglen, 1997). To explain this situation in PER, the open tags of the top ten most cited articles, total citation numbers and citation numbers per year are given in Table 2.

**Table 2.** The Ten Most Cited Articles between 2014 and 2021.

| Article | TC | ACY |
|---|---|---|
| Keller, J. M. (2016). Motivation, learning, and technology: Applying the ARCS-V motivation model. *Participatory Educational Research*, *3*(2), 1-15. | 129 | 21.83 |
| Caruth, G. D. (2014). Meeting the needs of older students in higher education. *Participatory Educational Research*, *1*(2), 21-35. | 78 | 9.75 |
| Solak, E., & Bayar, A. (2015). Current challenges in English language learning in Turkish EFL context. *Participatory Educational Research*, *2*(1), 106-115. | 75 | 10.71 |
| Caruth, G. D. (2018). Student engagement, retention, and motivation: Assessing academic success in today’s college students. *Participatory Educational Research*, *5*(1), 17-30. | 72 | 18.00 |
| Cakici, D. (2016). The use of ICT in teaching English as a foreign language. *Participatory Educational Research*, *4*(2), 73-77. | 66 | 11.00 |
| Caruth, G. (2014). Learning how to learn: A six-point model for increasing student engagement. *Participatory Educational Research*, *1*(2), 1-12. | 43 | 5.38 |
| Korkmaz, Ö., & Altun, H. (2014). Adapting computer programming self-efficacy scale and engineering students’ self-efficacy perceptions. *Participatory Educational Research*, *1*(1), 20-31. | 37 | 4.63 |
| Kayaoğlu, M. N., & Akbaş, R. D. (2016). An investigation into medical students' English language needs. *Participatory Educational Research*, *3*(4), 63-71. | 36 | 6.00 |
| Ekmekçi, E. (2016). Integrating Edmodo into foreign language classes as an assessment tool. *Participatory Educational Research*, *3*(4), 1-11. | 35 | 5.83 |
| Şahin, C. (2014). An analysis of the relationship between internet addiction and depression levels of high school students. *Participatory Educational Research*, *1*(2), 53-67. | 28 | 3.50 |

*TC= Total Citation, ACY= Average Citation per Year.*

As shown in Table 2, the ten most cited articles constitute approximately 2.88% of all articles, while the number of citations received by these articles corresponds to 35.76% of the number of citations received by the journal. Similarly, 10% of articles receive approximately 58.99% of citations. Meanwhile, taking account of this fact, to study the degree of skewness of the citation distributions in PER, the skewness of the citation distributions published by articles during 2014-2021 was analyzed using SPSS software. The descriptive statistics of the citation numbers were summarized as follows *minimum=0; maximum=129; skewness=6.247; kurtosis=50.231*. As expected, the positive skewed distributions were obtained, which means that the distribution of citations in PER is highly skewed to the right and has a long right tail. On the other hand, it is possible to see similar results in studies focused on citation skewness, especially educational journals and special education journals indexed in the WoS database between 2014 and 2018 (Orbay, Karamustafaoğlu, & Miranda, 2021; Örnek, Miranda, & Orbay, 2021).

### *Findings and Discussion for RQ3*

Most researchers who study a topic usually look at the title of the article to decide if it is relevant to them or scan it based on keywords related to research topics (Jamali & Nikzad, 2011). Therefore, the first impression that the title creates in the reader plays a major role in whether the article is read in detail or not. Therefore, the title is extremely important since it is the section that provides the most basic information about the content of an article (Hartley, 2008). A study of open-access journals has shown that the lower number of characters in the title is associated with the higher number of citations (Paiva, Lima, & Paiva, 2012). In a similar study, when articles published between 2007 and 2013 were examined, it was found that short-titled articles received a higher number of citations (Letchford, Moat, & Preis, 2015). The reasons for this relationship were listed as follows (Letchford, Moat, & Preis, 2015): journals with high impact factors limit the number of title characters; more recent research or research on emerging subjects have longer titles due to the need to be explained and they are published in less prestigious journals; short titles are easier to read, easier to understand, so attract more readers. Meanwhile, Elgendi (2019) used the machine learning approach to investigate the characteristics of highly cited papers and emphasized that a good title consists of 10±3 words (see: Elgendi (2019) for the recent detailed discussion).

Based on this data, the titles and citations of the articles published in PER were analyzed. It was determined that maximum (29) and minimum (4) words were used for the number of words in the title. The word count was sorted from less to high, and they were divided into three groups considering their median values: Less, Medium and High group, respectively, and the descriptive statistics for each group are given in Table 3.

**Table 3.** Descriptive Statistics of the Number of Words in the Title.

| | Group | N | M | Me | Min | Max | SD | AC |
|---|---|---|---|---|---|---|---|---|
| | Less | 123 | 9.03 | 9 | 4 | 11 | 1.689 | 6.53 |
| **The Number of Words in the Title** | Medium | 109 | 13.36 | 13 | 12 | 15 | 1.085 | 4.37 |
| | High | 115 | 19.09 | 19 | 16 | 29 | 2.805 | 3.02 |
| **Total** | | 347 | 13.72 | 13 | 4 | 29 | 4.625 | 4.69 |

*Note: M=Mean, Me= Median, Min=Minimum, Max=Maximum, SD=Standard Deviation, AC=Average Citations.*

As seen from Table 3, it was found that short-titled articles received more citations than long-titled articles (over 2 times greater). However, the citation difference among the groups is not statistically significant ($p>0.05$). It may be related to the limited number of articles. This result is agreement with the literature (Elgendi, 2019; Gnewuch & Wohlrabe, 2017).

A total of 1424 keywords were used in the articles. Word cloud work using *WordArt* program (2022) was carried out for these keywords and words that are frequently used in the title. In Figure 8, the size of each word is prepared according to the frequency of mentions in the title and keywords. Each word or group of words that appear in this image is fully compatible with the scope of the journal. The most repeated keywords are education ($f$=129), teacher ($f$=88), learning ($f$=79), science ($f$=57) and technology ($f$=53), while the words commonly used in the title section are teacher ($f$=85), student ($f$=82), education ($f$=77), learn ($f$=55) and school ($f$=55), respectively.

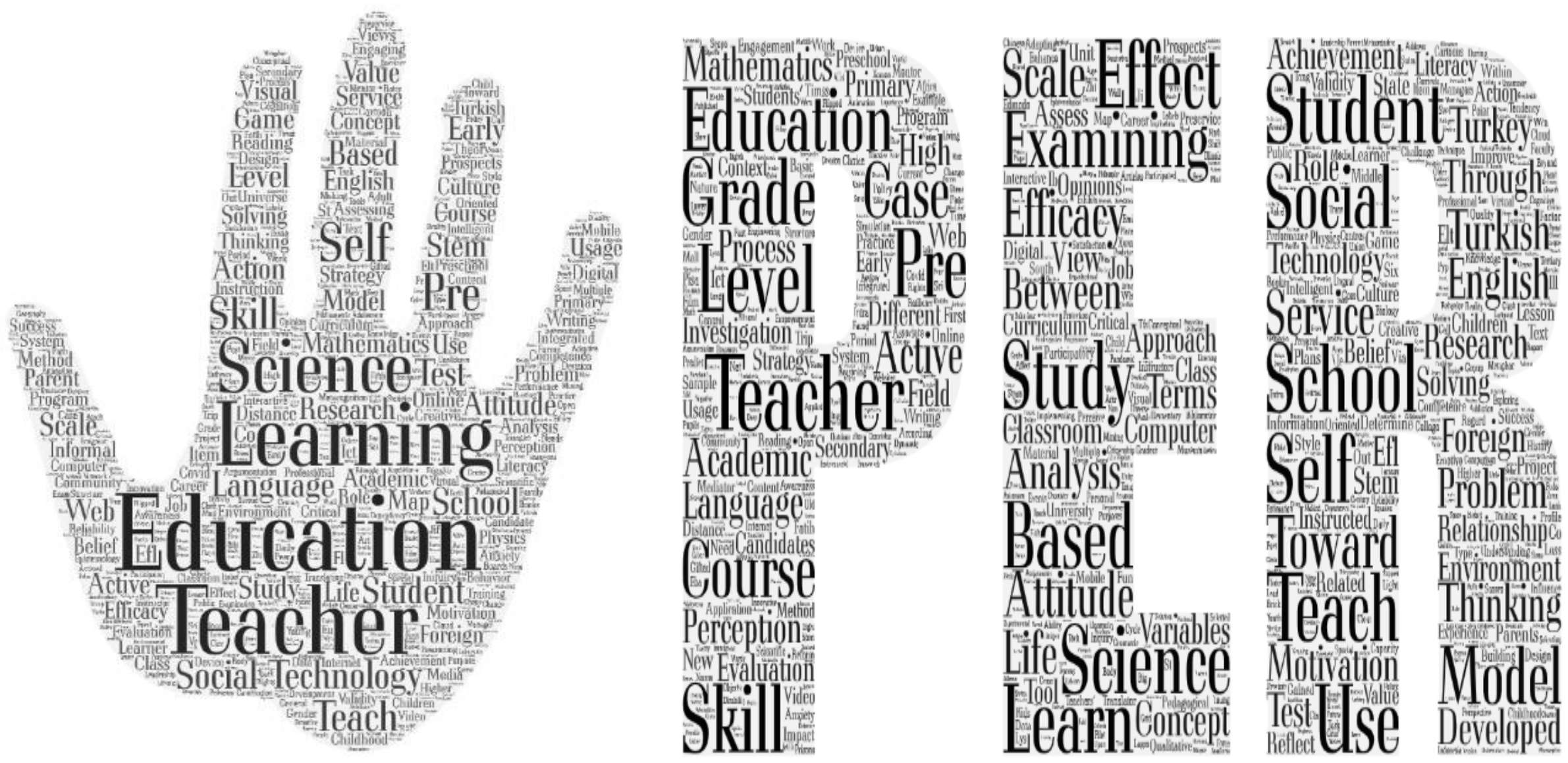

**a)** **b)**

**Figure 8.** Word Cloud for Keywords **(a)** and **(b)** Words in the Title.

On the other hand, when the keywords are carefully examined, some keywords are designated as very general concepts (sun, organ, word, etc.) or abbreviation words specific to that article (AWS, CIRC, IMMS, etc.) are used as keywords. Of course, keywords should show the general topic, but they should not be kept too broad and should not consist of acronyms specific to that article.

### *Findings and Discussion for RQ4*

The open access publication of articles that play important roles in the construction, dissemination and use of scientific knowledge presents an important opportunity especially for researchers with limited access to articles. On the other hand, open access articles are naturally read more and can receive more citations as a result (Piwowar et al., 2018; Piwowar, Priem, & Orr, 2019). There was a statistically significant positive correlation ($r_S=0.289$ and $r_P=0.524$; $p<0.01$) between the number of downloads of articles and the number of citations they received. On the other hand, there were no correlations among the number of citations and other bibliometric indicators such as the number of authors, the number of words in the title.

In a similar study on literature medical journals (Xue-li, Hong-ling, & Mei-ying, 2011), this correlation was found at $r=0.491$. In another study (Nieder, Dalhaug, & Aandahl, 2013), the correlation level was limited (from $r=0.1$ to $r=0.53$) depending on the type of documents. On the other hand, a study of publications on two information systems journals between 2002 and 2012 found a positive high correlation between the number of downloads and citations at $r=0.77$-$0.76$ (Schlögl et al., 2014). In a recent similar study focus on Library Philosophy and Practice journal, Arslan et al. (2022) found out that the correlations between download and citation counts from Scopus and Google Scholar databases were determined to be statistically significant positive ($r_S=0.261$ and $r_P=0.310$; $r_S=0.636$ and $r_P=0.356$; $p<0.01$), respectively. Meanwhile, it is possible to see a broad critique of the analysis of the relationship between download and the citation numbers by Hu et al. (2021).

PER, which has been published with open access since 2014, continues to live its publishing life with open access via DergiPark platform since 2019. Meanwhile, it should be note that the descriptive statistics of the number of downloads were summarized as follows *mean=304.19, minimum=81; maximum=2000; skewness=3.451; kurtosis=18.063*. As expected, the positive skewed distributions were obtained, which means that the distribution of the number of downloads in PER is highly skewed to the right and has a long right tail. When the number of downloads data was analyzed by dividing them into two separate groups, before and after joining the DergiPark, it was observed that there was a significant difference between the groups. As clearly seen from the mean rank values in Table 4, this difference is in favor of the group after DergiPark.

**Table 4.** Mann Whitney U Test Results for the Number of Downloads.

| | **Group** | **N** | **Mean Rank** | **Sum of Ranks** | **U** | **z** | ***p*** |
|---|---|---|---|---|---|---|---|
| **The Number of Downloads** | Before DergiPark | 178 | 103.96 | 18504 | 2573 | -13.349 | *.0000* |
| | After DergiPark | 169 | 247.78 | 41874 | | | |

On the other hand, the change in the average number of citations per article and median citations over the years were summarized in Figure 9. It was observed that the average number of citations per article had exceeded the general trend line. Finally, according to these findings, PER had gained significant momentum in academic standards and visibility since the umbrella of joining DergiPark in 2019.

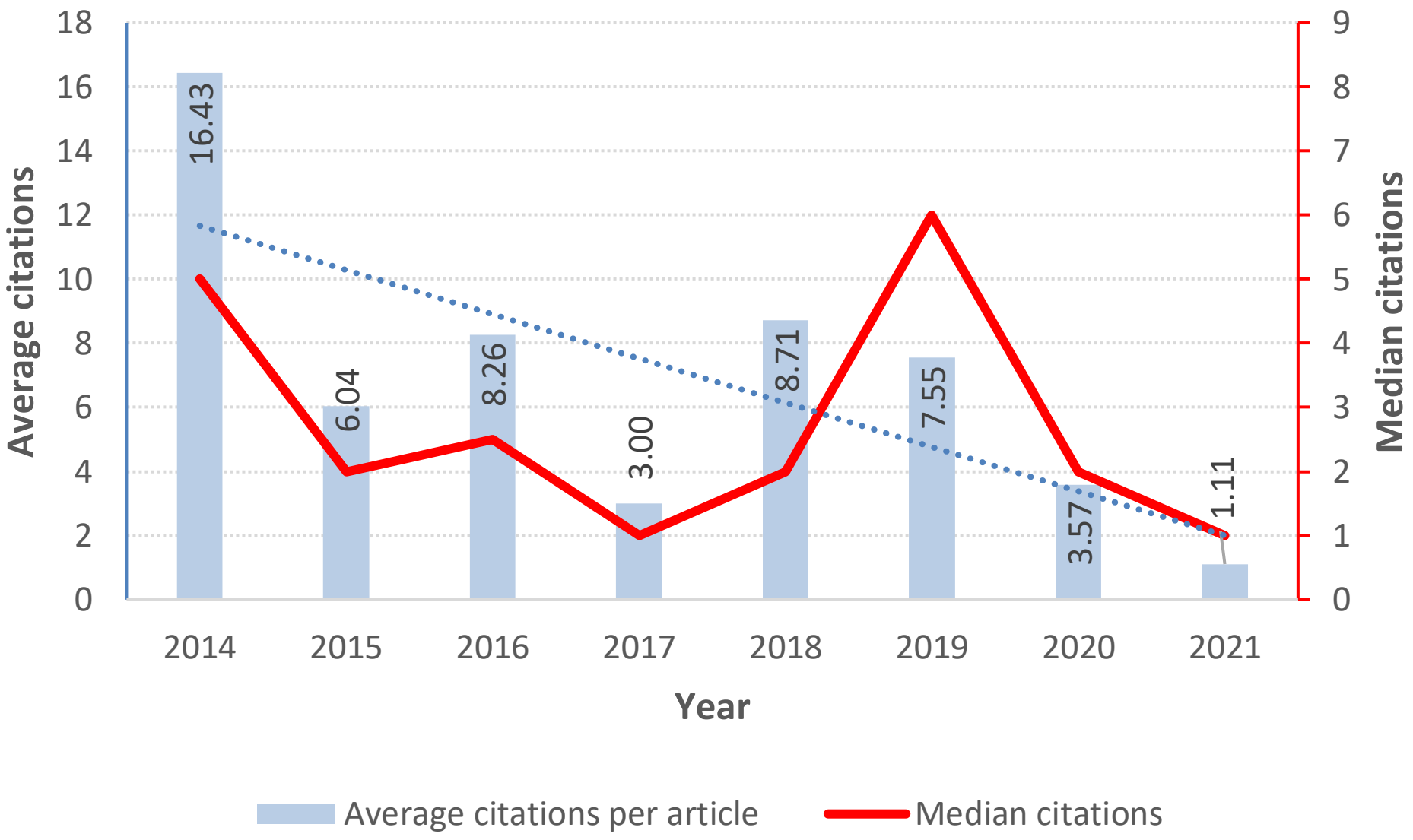

**Figure 9.** Year-wise Distribution of Average Citations per Article and Median Citations.

## Conclusion

Academic journals play an indispensable role among the official communication languages of science in the construction, dissemination, and use of information. In this context, PER has made a significant distance towards internationalization by contributing to the field of education and soon starting to be indexed in important databases such as ERIC and Scopus. It was concluded that 83.72% of the authors were in Turkey, there was no “institutional localization” among the institutions contributing at the national level and that they had achieved significant success in terms of national recognition. However, it is seen that there are many areas open to development at the point of international recognition. What to do at this point is summarized below.

- The fact that researchers from Anglo-Saxon and Continental European countries, especially USA, which stands out in the field of education with both article productivity and the widespread influence of articles (Sezgin, Orbay, & Orbay, 2022a; Sezgin, Orbay, & Orbay, 2022b), has been limited in PER is an important issue that must be overcome in front of the internationalization of the journal. If it is taken into account that there are enough editorial board members from these countries in this regard, the more active participation of the board members and their affiliated institutions with their articles will provide significant gains to the journal. On the other hand, as Allik et al. (2020) emphasized, low-population, well-managed countries with long democratic backgrounds (Netherlands, Belgium, Finland, etc.) should be encouraged to participate in the journal in their researchers from these countries, based on the fact that they are very successful in transforming their economic success into high-quality research in the field of education as well as in all fields.
- To increase the readability and the visibility of the articles and to prevent possible mistakes by the authors, the preparation of some guidelines regarding the fundamental concepts of writing a research paper such as the determination of the title, construction of the abstract, choosing the right keywords etc. for the researchers submitting manuscripts to the journal are of great importance from the writing strategies point of view (Gastel & Day, 2017); in addition, it would be appreciated as an important step to share these guidelines in the journal web page to improve the publishing standards of PER. Furthermore, many journals with a high international reputation utilize the statistics of past years such as the acceptance rate, the average or median time to first decision, and average or median review time to inform the researchers. Application of a similar procedure in PER will be beneficial, fruitful, and sound to provide necessary information for the researchers planning to submit manuscripts to the journal for a possible evaluation for publication.
- It is seen that only one review paper was published during the period of bibliometric analysis. Especially in the field of education, the original studies are scanned in great detail and widely, and the systematic reviews in which the findings are synthesized and the meta-analysis reviews where the results of the studies made by different researchers on the same subject are examined with appropriate statistics in quantity and quality are very important in determining the direction of the researchers. Although the review articles will give important clues to the researchers who will start researching on a new topic, they will contribute to the recognition of the journal and the citation issue, which is an important measure of the widespread impact of the paper (Miranda & Garcia-Carpintero, 2018). In this context, it will be an important step to call for articles by inviting researchers who come to the forefront with their national and international publications in their fields.
- With the growth of scientific literature and the widespread use of the internet, social media platforms such as Researchgate, Twitter, Linkedln, Facebook, and blogs are becoming more and more important in communicating the results of scientific research to large audiences (Shrivastava & Mahajan, 2022; Sudah et al., 2022). Considering these latest developments, interactive applications such as sharing articles and prominent findings in articles and interaction between researchers and authors can be developed in social media environments, which will be established in PER to increase the positive meaningful relationship between the number of downloads of articles and the number of citations they receive. Similarly, it would be useful to switch to “Highlights”, which consists of three to five items that have recently been implemented in journals published by many internationally recognized publishers (Elsevier, Taylor & Francis, John Wiley & Sons, etc.), featured in articles and helped to increase their discovery in search engines.

Finally, the findings can guide authors during the writing phase of studies and the editors and referees during the selection and evaluation phases. Therefore, what has been done and published in PER to date should be a new starting point to get to a better position.

## Limitations

Despite several notable contributions, this study had a few limitations. First, Bibliometric indicators based on citation number are time-dependent indicators and can change over time. Second, only the Google Scholar index was used in the citation search and self-citations were not checked in the study. Finally, the small number of articles constituting the sample is another important limitation. All these limitations reduce the generalizability of the results obtained in the study.

## Conflict of Interest

The authors declare that this study was conducted in the absence of any commercial or financial relationships that could be construed as a potential conflict of interest.

## Funding

This study received no specific grant from any funding agency in the public, commercial, or not-for-profit sectors.

## Acknowledgments

The authors would like to thank the anonymous referees for their useful comments.